# Simple and fast method for the analysis of $^{236}$U, $^{237}$Np, $^{239}$Pu and $^{240}$Pu from seawater samples by Accelerator Mass Spectrometry

*Mercedes López-Lora* [1,2]* – *Isabelle Levy*[3] – *Elena Chamizo* [1]

[1]Centro Nacional de Aceleradores (CNA). Universidad de Sevilla. Junta de Andalucía. Consejo Superior de Investigaciones Científicas. Parque científico y tecnológico Cartuja. Thomas Alva Edison 7, 41092, Sevilla, Spain

[2]Dpto. de Física Aplicada I, Escuela Politécnica Superior, Universidad de Sevilla, Virgen de África 7, 41011 Sevilla, Spain

[3]International Atomic Energy Agency, Environment Laboratories, MC 98000 Monaco

*Corresponding author:
E-mail address: mlopezlora@us.es

## Abstract

A simple and fast radiochemical procedure for the sequential extraction of U, Np and Pu from small-volume seawater samples (<10 L) is presented. The method has been developed and optimized for the final determination of $^{236}$U, $^{237}$Np, $^{239}$Pu and $^{240}$Pu by Accelerator Mass Spectrometry (AMS). It is based on the use of $^{242}$Pu as tracer for both Np and Pu isotopes. Samples are pre-concentrated by Fe(OH)$_2$ co-precipitation. TEVA® and UTEVA® resins are used in a very simplified way for the final purification of the Pu+Np and U fractions, respectively. The radiochemical yields of the three elements have been investigated in detail by alpha spectrometry (AS) and gamma spectrometry (GS). The obtained results indicate high and robust chemical yields for the three elements and similar ones for Pu and Np. Furthermore, the use of $^{242}$Pu as tracer for $^{237}$Np is validated by analyzing a reference seawater sample (IAEA-443) by radiometric techniques. We demonstrated that, if chemicals are properly chosen, processed blank levels can be kept at the same level of the extremely low detection limits that can be achieved by AMS ($10^5$-$10^6$ atoms/sample). The procedure is finally applied for the study by AMS of a reference seawater sample from the Mediterranean Sea (IAEA-418).

# 1. Introduction

$^{236}$U ($T_{1/2}$ = 2.34·10$^7$ y) and $^{237}$Np ($T_{1/2}$ = 2.14·10$^6$ y) are both considered as ideal oceanographic tracers because of their conservative behavior in seawater, long half-life and mainly anthropogenic origin. These radionuclides were introduced into the general environment on a global scale during the period of atmospheric nuclear weapons testing (1945-1980), being the $^{236}$U and $^{237}$Np estimated inventories 900-1100 kg and 1500 kg, respectively [1–3]. Locally, the accidental and/or controlled discharges from nuclear fuel reprocessing plants, NRP, are the most outstanding sources. E.g., it is estimated that about 85 kg and 170 kg of $^{236}$U and $^{237}$Np, respectively, have been entered the North Atlantic from different European NRPs (i.e. 60 ± 30 kg and about 23.5 kg of $^{236}$U have been released from Sellafield (U.K) and La Hague (France), respectively, while $^{237}$Np discharges are coming mainly from Sellafield) [4,5]. Furthermore, the combined study of $^{236}$U and $^{237}$Np might be a potential tool in oceanography to identify water-mass sources and transient times, as it has been the case of $^{236}$U and $^{129}$I [6]. Besides, the $^{237}$Np/$^{236}$U ratio can provide information on the contamination source in a specific environment. For instance, in surface soils samples solely influenced by global fallout, $^{237}$Np/$^{236}$U expected atomic ratios are in the 1.7 -2.5 range in the North Hemisphere (estimations from the $^{237}$Np/$^{239}$Pu and $^{236}$U/$^{239}$Pu ratios reported in [7] and [8], respectively). In contrast to this, the measured ratio from the IAEA-381 (i.e. seawater from the Irish Sea collected in 1993, impacted by the discharges from the Sellafield NRP) is 42 ± 3 [9], and ratios at the 0.001 level are expected for the debris released as a consequence of the Chernobyl accident (former Soviet Union, now Ukraine, 1986) [10]. In the last years, an increasing number of studies have been focused on $^{236}$U in the oceans because of its potential as oceanographic tracer [11–15]; however, despite their similarities, the number of $^{237}$Np results reported in seawater samples [4,16–19] are very scarce. The most likely reason is the complexity associated to the radiochemical procedures aimed at the Np purification.

On the other hand, $^{239}$Pu ($T_{1/2}$ = 2.41·10$^4$ y) and $^{240}$Pu ($T_{1/2}$ = 6.56·10$^3$ y), which are the most abundant Pu isotopes, present a very different geochemical behavior. Pu is a particle-reactive element and it is exported from the surface seawater with sinking particles which subsequently suffer remineralization processes. Similar to $^{236}$U and $^{237}$Np, Pu has been also introduced in the oceans mainly by global fallout and the operation of NRP. However, there exists a large database on the presence of Pu isotopes in the general environment, and the source-dependency of the $^{240}$Pu/$^{239}$Pu atom ratios has been widely documented. The minimum $^{240}$Pu/$^{239}$Pu atomic ratio, i.e. 0.013 ± 0.003, has been reported for the mineral *Trinitite,* a green-glassy mineral formed during the testing of the first US nuclear atomic weapon, Trinity (20kT, New Mexico, 1945) [20]; and the maximum ones correspond to areas influenced by the Chernobyl accident, with values of 0.57 and above [21].

The concentrations in the open ocean of $^{236}$U, $^{237}$Np and $^{239,240}$Pu can be extremely low. For instance, surface seawater from the Eurasian basin at the Arctic Ocean , impacted by the releases from European NRP, presents $^{236}$U concentrations in the (1-4)·10$^7$ atoms L$^{-1}$ range (i.e. from 8 to 30 nBq L$^{-1}$, and $^{236}$U/$^{238}$U atom ratios in the (1-4)·10$^{-9}$ range), and $^{237}$Np in the (1-10)·10$^7$ atoms L$^{-1}$ range (i.e. 0.1-1 µBq L$^{-1}$) [4,6]. Regarding Pu isotopes, $^{239+240}$Pu activities ranging from 1 to 20 µBq L$^{-1}$ (which means $^{239}$Pu and $^{240}$Pu concentrations at the level of 5·10$^5$-10$^7$ and 10$^5$-10$^6$ atoms L$^{-1}$, respectively) have been reported in deep-water columns from the Arctic Ocean [8]. Even lower concentrations are expected in other areas only influenced by the global fallout. For example, in the South Pacific Ocean, surface seawater samples present concentrations in the (4-7)·10$^6$ atoms L$^{-1}$ or 4-6 nBq L$^{-1}$ range for $^{236}$U (i.e. $^{236}$U/$^{238}$U atom ratios in the (4-7)·10$^{-9}$ range) [22] and in the 0.5-1.4 µBq L$^{-1}$ range for $^{239+240}$Pu activities (which means concentrations at the level of 10$^5$ and 10$^4$-10$^5$ atoms L$^{-1}$ for $^{239}$Pu and $^{240}$Pu, respectively) [23]. Although there is no information about $^{237}$Np in this area, its concentrations are expected to be also very low, at the same level of $^{236}$U.

The study of these radionuclides at environmental levels by using small-volume seawater samples (i.e. below 10L) has recently become possible as a result of the high sensitivity reached by modern Mass Spectrometry (MS) systems. Low detections limits for $^{237}$Np and $^{239,240}$Pu analysis are offered by Thermal Ionization Mass Spectrometry (TIMS) and High-Resolution Inductively-Coupled-Plasma Mass Spectrometry (HR-ICP-MS) [24–26]. However, an effective U chemical separation is required in order to achieve the necessary purification levels [26] due to the presence of the naturally-occurring U isotopes (mainly $^{238}$U and $^{235}$U) results in the most important limitation for $^{237}$Np and $^{239}$Pu analysis. The presence of these U isotopes increase considerable the final backgrounds not only by tailing effect in the relative close mass of interest but also through the formation of molecular isobars (e.g. $^{238}$U$^{1}$H$^{+}$ in the case of $^{239}$Pu) [26]. Therefore, the introduction of additional purification steps becomes indispensable, becoming the radiochemical procedures more complex and time consuming. These problems caused by U molecular isobars can be overcome by using Accelerator Mass Spectrometry (AMS). This technique reaches the highest sensitivity mainly thanks to the destruction of the molecules in the stripping process, with occurs in the terminal of an electrostatic tandem accelerator. Hence, the use of AMS allows the simplification of the radiochemical procedures as U purification levels required are lower than for other MS techniques. Moreover, regarding $^{236}$U analysis, AMS also offer the highest sensitivities tanks to the destruction of the molecular isobar $^{235}$U$^{1}$H [27].

In the recent years, many efforts have been focused on the development of methods for the sequential extraction of different actinides from the same seawater aliquot, given the high value and uniqueness of the samples involved in oceanographic expeditions [28–32]. Methods aimed at minimizing the sample volume are required in order to simplify the logistics of the sampling campaigns, and increase the number of samples. Usually, actinides are first pre-concentrated from the sample volume and the complexity of this step depends on the sample size (e.g. different co-precipitation stages are needed for large-volume samples, above 100 L, for $^{239+240}$Pu and $^{237}$Np determinations by alpha-spectrometry (AS)). Then, actinides are purified using ion chromatography methods [33]. In the last decade, TEVA® and UTEVA® resins (Triskem Industries, Inc.), which are specifically designed for actinides, have demonstrated to be the most convenient ones because of their ease of use [34].

On the other hand, most of the reported methods are not focused on the extraction of $^{237}$Np mainly due to the lack of a long-life Np isotope available to be used as a yield-tracer. The only choice would be $^{236}$Np (T$_{1/2}$ = 1.54·10$^5$ y), but the stock solutions that are currently commercially available contain significant traces of $^{237}$Np. As for the $^{236}$U AMS analysis, its use won´t be convenient, since $^{236}$Np and $^{236}$U are isobars. $^{239}$Np (T$_{1/2}$ = 2.35 d) has been also used as a yield-tracer for $^{237}$Np determinations [18,35], but due to its short half-life, it has important disadvantages: i) $^{239}$Np has to be previously milked from a parent $^{243}$Am solution [32]; ii) that parent solution might contain traces of $^{241}$Am, which subsequently decays to $^{237}$Np producing contamination; iii) $^{239}$Np recovery in each step has to be monitored by gamma spectrometry (GS); and iv) the whole processing of the samples has to be completed in a few days because of the fast decay of the tracer. To overcome all these problems, some authors have pointed out the possibility of using $^{242}$Pu (T$_{1/2}$ = 3.73·10$^5$ y) instead, since both Pu and Np fractions can follow the same chemical behavior if their oxidations states are properly controlled [30,31,36]. In any case, most of these methods have been developed for conventional MS techniques, and complex and time-consuming extraction chromatography procedures are needed in order to ensure a reliable purification of the different fractions [19,37].

In this work, we present a simple and fast sequential method to separate U, Np and Pu from small-volume seawater samples for the final analysis of $^{236}$U, $^{237}$Np, $^{239}$Pu and $^{240}$Pu by AMS. The method is based on the use of $^{242}$Pu as the yield tracer of both $^{237}$Np and Pu isotopes, which are recovered in the same fraction (from now on, Pu+Np fraction), and is an extension of the procedure previously reported by our group for the sequential extraction of U and Pu [29]. Samples are pre-concentrated by only one Fe(OH)$_2$ co-precipitation step and TEVA® and UTEVA® resins are used in a very simplified way for the final purification of the U and Pu+Np fractions. Radiometric techniques (AS and GS) have been used to assess and optimize the performance of the method, and for its final validation by analyzing a reference sample provided by the International Atomic Energy Agency (IAEA) (i.e. IAEA-443, Irish Sea seawater). The method is finally applied for the analysis of the IAEA-418 (Mediterranean seawater sample) on the 1 MV AMS system at the CNA following the recently developed technique [38]. Laboratory background levels have been also optimized in order to take advantage of the low detection limits of the AMS technique for the study of environmental samples. A description of the method and its performance, and the different validation exercises carried out are described in the following sections.

## 2. Experimental

### 2.1 Reagents, materials and spikes

During the sample processing, M-RO water (i.e. Milli-RO® water, pure distilled water) and acids of the highest purity were used in every stage to prevent contamination in the AMS analysis. The different reagents used throughout the procedure were supplied by Panreac, Honeywell Fluka, Acros Organics and Labosi. Besides, all these reagents were studied in order to evaluate possible sources of contamination (see section 3.4).

Ion chromatography separations were performed using TEVA®, UTEVA® and Pre-filter resins (2 mL columns of 100-150 µm particle size; Triskem Industries, Inc.).

For the AMS cathode preparation step, a Fe(III) solution provided by *High Purity Standards* (HPS, England) was used, with a certified $^{238}$U concentration below the 0.5 ppb level. The Nb powder was provided by Sigma-Aldrich® (purity of 99.8%).

During the testing of the procedure by radiometric techniques, the following spikes were used: $^{232}$U ($T_{1/2}$ = 68.9 y), provided by National Physical Laboratory (NPL, England); and $^{239}$Np ($T_{1/2}$ = 2.35 d), milked from a $^{243}$Am ($T_{1/2}$ = 7.37·10$^3$ y) solution provided by Harwell Laboratory (Atomic Energy Research Establishment, U.K.). $^{239}$Np, in secular equilibrium with its $^{243}$Am parent, was separated and purified according to the method described in [32,39].

For the $^{236}$U, $^{237}$Np and $^{239,240}$Pu AMS measurements, the following spikes were used: $^{233}$U ($T_{1/2}$ = 1.59·10$^5$ y), provided by CEA (French Alternative Energies and Atomic Energy Commission), and $^{242}$Pu ($T_{1/2}$ = 3.73·10$^5$ y), provided by NPL.

**2.2 Samples**

Two sets of samples were used for method testing purposes: i) seawater samples collected at the Monaco coast (2.5 and 10 L aliquots), and ii) some mL of diluted $HNO_3$ spiked with $^{232}U$, $^{239}Np$ and $^{242}Pu$ (i.e. called in-house column-blank samples).

For the validation of the proposed procedure, two IAEA reference seawater samples were used: IAEA-443, Irish Sea seawater influenced by Sellafield NRP [40]; and IAEA-418, surface seawater collected in 2011 at the DYFAMED station, in the north-western Mediterranean Sea [41]. $^{239}Pu$ and $^{240}Pu$ reported activities from the IAEA-443 are (8.2 ± 0.8) mBq kg$^{-1}$ and (7.0 ± 0.6) mBq kg$^{-1}$, respectively, and $^{238}Pu$ certified activity is (3.1 ± 0.1) mBq kg$^{-1}$[40]. IAEA-443 aliquots come from the same original sample than IAEA-381 aliquots [40], which certified $^{237}Np$ activity is (8.7 ± 0.5) mBq kg$^{-1}$ [9]. All these uncertainties are expressed at the k=2 (95%) confidence level. The IAEA-418 is only certified for $^{129}I$. However, based on the reported information on samples collected at the same station, it can be expected $^{237}Np$ and $^{239+240}Pu$ activities of (0.223 ± 0.071) µBq kg$^{-1}$ and (14 ± 2) µBq kg$^{-1}$, respectively [18,42].

Every seawater sample was filtered with a 0.45 µm pore-size filter to remove coarse particles before applying the chemical procedure.

**2.3 Techniques**

AS determinations were performed in the IAEA Environment Laboratories (IAEA-NAEL, Monaco) by using low background ion-implanted silicon detectors (Ultra-AS Ortec) with an active area of 450 mm$^2$ according to [43]. For the evaluation of the spectra, Alphavision 6 software (Ortec) was used.

$^{239}Np$ analysis by GS were developed at IAEA-NAEL following the method described in [43]. The measurements were performed by using a High Purity Ge detector with 44% of relative efficiency (Eurisys); and, for the evaluation of the spectra, Gammavision 6 software (Ortec) was used.

AMS determinations were performed on the 1 MV AMS system at the CNA. Details about the current status of the facility are given in [44,45]. The setup for the $^{236}$U measurements is discussed in [46,47]. $^{237}$Np and Pu isotopes are analyzed jointly from the same cathode (i.e. the final sample adapted to the optimal matrix for the actinides extraction in the ion-source) following the recently developed techniques detailed in [38].

All uncertainties presented in this article are expressed at the k = 1 (68%) confidence level, except where specified otherwise.

**2.4 Proposed radiochemical method for $^{236}$U, $^{237}$Np and $^{239,240}$Pu determinations**

The proposed method is schematized in Fig. 1 and explained in detail below.

*2.4.1 Sample pre-treatment*

Filtered seawater samples are acidified by adding concentrated HCl until pH 2. About 3 pg of $^{233}$U and $^{242}$Pu are added as spikes to control the radiochemical yield of U and Np+Pu fractions, respectively, and to quantify their concentrations in the sample matrix following the isotope-dilution method. They are also used as the reference isotopes during the corresponding AMS measurement of $^{236}$U and $^{237}$Np+$^{239,240}$Pu [38].

*2.4.2 Co-precipitation of the actinides with Fe(OH)$_2$*

Actinides are pre-concentrated from the bulk sample by Fe(OH)$_2$ co-precipitation following different steps: (i) potassium metabisulphite (K$_2$S$_2$O$_5$) is added to the samples; (ii) once dissolved, Fe(II) is added as iron sulphate (FeSO$_4$) salt and mixed; and (iii) finally, the Fe(OH)$_2$ precipitate (i.e. blue-greenly particles, see Fig. 2) is obtained by increasing the pH to 9 with concentrated ammonia (NH$_4$OH). For 10 L samples, 5 g of K$_2$S$_2$O$_5$ and 500 mg of Fe(II) are used, which means that 0.5 g of K$_2$S$_2$O$_5$ and 50 mg of Fe(II) are added for each liter of the processed sample. In the case of seawater aliquots below 5 L, these concentrations are increased in order to ensure that co-precipitation process works properly: 2.5 g of K$_2$S$_2$O$_5$ and 250 mg of Fe(II) are used for 2.5 L samples, and 1.25 g of K$_2$S$_2$O$_5$ and 125 mg of Fe(II) are used for 1 L samples. After about 1-2 hours of settling of the Fe(OH)$_2$ particles the supernatant is removed. The Fe(OH)$_2$ precipitate is transferred to 500 mL plastic bottles, centrifuged at 3000 rpm during 15 min, and the resulting supernatant is discarded. The precipitate is dissolved with concentrated HNO$_3$, transferred to a 250 mL glass beaker, evaporate to dryness, and treated with concentrated HNO$_3$ several times to eliminate HCl traces (i.e. until the vanishing of the orange fumes).

### *2.4.3 Chromatography separation*

Purification of U and Np+Pu is carried out using two extraction chromatography resins, TEVA® and UTEVA®, placed in tandem. A fast paper filter (14-18 µm) is arranged in a funnel on top of the TEVA® column to retain any particle which could obstruct the resins. Resins are conditioned by eluting 20 ml of 3M HNO$_3$, which is the matrix where Pu and Np exhibits a high retention on TEVA® columns and U on the UTEVA® at the same time [34,48].

Prior to the ion exchange separation, the samples are treated to ensure a high and equal uptake of Pu and Np in the TEVA® resin, by adjustment of their oxidation states to (IV). This is achieved in two stages: i) the precipitate is dissolved in 18 mL of 1 M HNO$_3$ and 200 mg of Mohr's Salt, i.e. (NH$_4$)$_2$Fe(SO$_4$)$_2$·6H$_2$O, is added to reduce Pu to Pu(III) and Np to Np(IV); ii) Pu is oxidized to Pu(IV) adding 3 mL of concentrated HNO$_3$ (70%). The load solution (21 ml of 3 M HNO$_3$) is then added to the columns and, once it has been eluted, 5 mL of 3 M HNO$_3$ used to rinse the beaker are added to the column, the filter is rinsed with another 5 mL of 3 M HNO$_3$ and an additional rinse of 20 mL is finally added. Then both resins are separated in order to elute Pu+Np from the TEVA® and U from the UTEVA®. Pu+Np are separated as it follows: 20 mL of 10 M HCl are added to the TEVA® to remove Th and, prior to eluting Pu and Np from the resin, a 2 mL Prefilter resin is placed below the TEVA® column to avoid any traces of organic matter in the final purified solution. The Prefilter has to be previously conditioned by adding 15 ml of 0.1 M HF – 0.1 M HNO$_3$. Finally, Pu+Np are eluted from the TEVA® with 25 mL of 0.1M HF – 0.1 M HNO$_3$. U is eluted from the UTEVA® with 15 mL of 0.1 M HNO$_3$ and 10 mL of distilled water.

### *2.4.4 Preparation of the AMS cathodes*

To prepare the AMS sources, every sample needs to be dispersed in an iron oxide matrix mixed with Nb powder, which is the most optimum medium for the extraction of the corresponding actinide as $AnO^-$ in the Cs-sputtering ion-source [44,45]. Two different methods are used for U and for Np+Pu fractions. (i) For the uranium fractions, a conventional method was applied [8]: U is co-precipitated with $Fe(OH)_3$ after adding 1 mg of Fe(III) and increasing the pH with ammonia to 8-9. The precipitated is washed with 1 mL of ethanol to remove the organics coming from the resin and transferred to a 2 mL quartz crucible. The precipitate is dried on a hot plate at about 50°C, then oxidized at 650°C during 1 hour in a muffle furnace and, finally, mixed with 3 mg of Nb powder and pressed into an appropriated aluminum cathode. (ii) In the case of Pu+Np samples, a different protocol was adopted to prevent chemical fractionation between the two elements. The final solution is recovered from the TEVA® in a 25 ml centrifuge Teflon® tube. Then, 1 mg of Fe(III) is added and the sample is dried on a hot plate. No organic remains are expected to be in the final dried samples as we are using a prefilter after the TEVA®. The dried iron is transferred to a 5 mL ceramic crucible for the final oxidation at 650°C during 1 hour. The resulting sample is mixed with 3 mg of Nb powder and pressed into the aluminum cathode.

**2.5 Testing of the proposed procedure by radiometric techniques**

In this work, radiometric techniques were used with the aim of studying the performance of the proposed method, allowing a direct quantification of the U, Np and Pu radiochemical yields in different stages.

*2.5.1 Estimation of U and Pu radiochemical yields by AS*

Uranium and Plutonium recoveries were quantified by AS by studying the activity of a well-known initial spike. Pu results were obtained by using $^{242}Pu$ as yield tracer. To this end, about 20 mBq (or 130 pg) of $^{242}Pu$ were added to the testing samples (i.e. Monaco seawater) in the pre-treatment step. As for U, the naturally present $^{238}U$, whose initial concentration in these samples was (3.2 ± 0.1) µg $L^{-1}$, was used as a reference. Additionally, in order to study the achieved U recovery in the pre-concentration step independently (section 2.4.2), samples were spiked with $^{232}$U (about 13 mBq or 35 pg) after the Fe(OH)$_2$ co-precipitation stage. That stage was demonstrated in the past to be kay as long as U recovery is concerned [29].

The performance of the extraction chromatography step was evaluated in an independent experiment by analyzing 6 column-blank samples (i.e. nitric acid spiked with $^{232}$U, $^{239}$Np and $^{242}$Pu as explained in section 2.2). Slight modifications in the acid used in the oxidation state adjustment procedure were introducing for testing its solidity: i) one set of samples (Blk-1, Blk-2, Blk-3) was treated with 17 mL of 1M HNO$_3$, 200 mg of Mohr's Salt and 2.75 mL of HNO$_3$ (70%); and ii) a second one (Blk-4, Blk-5, Blk-6) with 20 mL of 1M HNO$_3$, 200 mg of Mohr's Salt and 3.25 mL of HNO$_3$ (70%) (see section 2.4.3 for details).

In every case, the AS sources were prepared from the U and Pu final solutions following the electrodeposition method described in [49]. In order to control the losses in that additional stage, every sample was treated twice (i.e. the supernatant in the electrodeposition cell after a first electrodeposition, was processed once again following the same method). The AS sources from the second electrodepositions were counted to ensure that non-significant activity was detected. For each AS detector, the efficiency was studied by using AS standard sources containing well-known activities of $^{239}$Pu, $^{241}$Am and $^{244}$Cm.

### *2.5.2 Estimation of Np radiochemical yields by GS*

Neptunium radiochemical yields were studied in each stage independently by GS by using $^{239}$Np as the yield tracer. To this end, the resulting solutions were transferred to appropriate containers and measured on-site. $^{239}$Np activities were evaluated in each step taking into account the losses due to their radioactive decay.

### *2.5.3 Validation of the procedure: measurement of IAEA-443 by radiometric techniques*

A first feasibility study of using Pu as a yield tracer for Np was developed by using AS and GS for the analysis of the IAEA-443 reference material, being currently the only available seawater sample with certified values for $^{237}$Np to the best knowledge of the authors. An inconvenience appears for the study of $^{237}$Np by AS since the $^{242}$Pu alpha emissions overlap with the ones of $^{237}$Np (i.e. 4.9005 MeV and 4.7880 MeV, respectively), preventing its use as a yield tracer for its measurement by AS. Therefore, an alternative approach was adopted this time: $^{238}$Pu, whose concentration is also certified in this sample, was used as the reference nuclide instead. The samples were also spiked with $^{239}$Np and its yield controlled by GS, to get a direct estimation of the Np yield.

3 subsamples were prepared by diluting 1 L of the original reference sample with 9 L of Monaco seawater, to simulate the typical volumes used for the AMS analysis of general seawater samples, and processed following the proposed procedure. Given the high $^{237}$Np concentration in the original sample (i.e. (8.7 ± 0.5) mBq kg$^{-1}$), the resulting sample was still compatible with AS determinations. The contribution of the Monaco seawater aliquots to the final $^{237}$Np activity of the samples was considered negligible.

## 3. Results and discussion

### 3.1 Pre-concentration step: Fe(OH)$_2$ co-precipitation

In previous works, it was demonstrated that the performance of the Fe (III)-based co-precipitation step (i.e. Fe(OH)$_3$) is critical as for the efficient U pre-concentration and it was evidenced a dependence of the U recovery yield with the amount of Fe added as a carrier [29,50]. Moreover, previous studies pointed out the potential of using the Fe(II) co-precipitation for the pre-concentration of actinides [51], although complex precipitation procedures involving several steps were performed in this work. In order to study and optimize the pre-concentration step of the proposed method, based on Fe(II) (i.e. Fe(OH)$_2$), U and Np recoveries were quantified separately as it is described in the sections 2.5.1 and 2.5.2 (i.e. from the naturally present $^{238}$U in the test sample and addition of $^{232}$U prior to the column separation by AS, and from the added $^{239}$Np spike by GS, respectively). The effect of the pH, which has revealed to be critical, was evaluated by performing two different tests (Table 1). In a first experiment, the precipitates were obtained by increasing the pH to 8-8.5 following the procedure described in section 2.4.2 (samples A, B and C). They presented a characteristic green color and the supernatant was murky, indicating the presence of Fe(OH)$_2$ particles in solution. Those first precipitates (1$^{st}$ ppt) were recovered and the corresponding Np yields were quantified by GS. The supernatants were mixed again, let settle for 2-3 additional hours and the resulting precipitates were recovered once again. Np yields were quantified for these second precipitates (2$^{nd}$ ppt). Finally, both precipitates were combined, spiked with $^{232}$U, and processed according to the proposed procedure in order to obtain the U recoveries by AS as it is explained in section 2.5.1. In a second experiment (samples D, E and F), the co-precipitation was performed by increasing the pH to 9-9.5. In this case, the first precipitates (1$^{st}$ ppt) presented a dark bluish green color and the supernatant was clearer than the previous experiment. Therefore, it was assumed that a second recovery of the precipitate (2$^{nd}$ ppt) was not necessary this time.

Neptunium recoveries obtained for the samples at pH 8.5 in the first precipitate were lower than 70% (i.e. average recovery of 68% and 1% of standard deviation (SD)), but they were compensated by the second precipitated (average recovery of 13% and 3% of SD), resulting in a final average recovery of 82% (3% SD). The results obtained from the samples at pH 9-9.5 were higher despite the fact that only one precipitate was recovered. All the values were higher than 85%, with an average recovery of 93% (4% SD).

Uranium recoveries were below 80% for the three samples at pH 8-8.5, being especially low in B sample, i.e. (25.2 ± 3.8)%. The average recovery in this case was 60% (30% SD). U yield results increased in the case of the pH 9-9.5 samples, being the co-precipitation yields above 80% in every case, with an average value of 98% (9% SD). About the same value was obtained in the other processed samples in this work (18 samples overall), which validates the solidity of this result. In contrast, Fe(III) co-precipitation, which is one of the most common pre-concentration methods used for actinides, gives low U recoveries when sample volumes larger than 5 L are processed using a similar procedure. In that case, it becomes mandatory the use of larger amounts of Fe(III) carrier in order to collect the uranium efficiently, making the method more complex [29,50].

## 3.2 Extraction chromatography step: oxidation state adjustment

The Np and Pu oxidation-states adjustment performed before the extraction chromatography step is a critical point of the proposed method. As both elements are eluted from the TEVA® resin in the same step, it is essential for the use of $^{242}$Pu as the yield tracer of both elements to obtain the same chemical recoveries. This means that the same fraction of both elements must be in (IV) oxidation state, which is the one retained by the TEVA® resin. Moreover, such adjustment must be efficient in both cases, in order to get good chemical yields in this step. Note that the oxidation-sate adjustment is not critical in the case of U as long as the final yield is concerned, since its most likely oxidation state is U(VI), which is the one retained by the UTEVA® resin.

As it is described in the section 2.3.3, following the reduction of Pu and Np to Pu(III) and Np(IV) respectively, concentrated $HNO_3$ is added to oxidize Pu to Pu(IV). The volume of $HNO_3$ has to be high enough to ensure the complete oxidation of Pu, but keeping the Np as Np(IV) (i.e. too much acid might further oxidize Np(IV) to Np(V), which is not retained by the TEVA® resin [52]). In order to test the soundness of this step, Np, Pu and U recoveries solely from the extraction chromatography step were studied by using 2 sets of column-blank samples (see section 2.2), which were treated with slightly different amounts of nitric acid (see section 2.5.1). The recoveries obtained for these samples are shown in the Fig. 3 and Table 2. Results are high in all the cases and very similar for the two sets of samples (i.e. U recovery for the Blk-5 sample was the only value under 80%). In all the cases, Np and Pu recoveries agree within uncertainties. The Np/Pu recoveries ratios are 1.03 ± 0.06 and 0.9 ± 0.1 for the first and second sets of samples, respectively. These results show that variations in the added volume of $HNO_3$ of less than 0.5 mL, do not affect the Np and Pu recovery yields in the extraction chromatography step.

## 3.3 Overall U, Np and Pu radiochemical yields

The overall U, Np and Pu radiochemical yields for the matrixes of interest, were obtained by processing 2.5 L and 10 L of seawater following the proposed procedure (section 2.4). The obtained results are shown in Table 3 and Fig. 4.

Regarding the U results, the recoveries were high (> 80%) in all cases, both in the 2.5 L and in the 10 L aliquots. The average recovery was 93% (8% SD). Np and Pu recoveries were similar and over 80% in all the samples except for Sample 2, which presents lower recoveries but also similars for both Np and Pu. The average recoveries were (80 ± 10)% for Np and (80 ± 10)% for Pu. Np/Pu ratios from the obtained radiochemical yields for each sample are shown in the last column of Table 3. The average value for this Np/Pu yield ratios is 0.98 with a standard deviation of 0.04.

**3.4 Background optimization**

AMS offers detection limits at the $10^5$-$10^6$ atoms/sample level for every actinide. These levels are similar to the $^{236}$U, $^{237}$Np and $^{239,240}$Pu concentrations in general seawater samples so, in principle, their measurement using small-volume aliquots (i.e. below 10 L) should be possible. In practice, the achievement of such background levels entails a careful control of the reagents used throughout the chemical procedure.

In order to study the background introduced by the reagents used in the proposed procedure, different tests were conducted. Negligible amounts of $^{237}$Np and Pu isotopes were detected in every studied reagent, with processed blank levels below $10^5$ atoms. However, significant traces of $^{236}$U were identified in one of the three studied Fe(II) salts used as a carrier in the first co-precipitation stage, at the level of $10^7$ atoms·g$^{-1}$. This level is too high to process environmental samples, which could present $^{236}$U concentrations at the level of $10^6$ atoms·L$^{-1}$ or even lower. Therefore, it is strongly recommended to investigate the $^{236}$U background coming from the Fe(II) salt whenever the procedure is applied for the first time in a laboratory or when a new one is used, even if it comes from the same supplier.

**4.4 Validation of the proposed procedure by radiometric techniques: IAEA-443**

The necessary validation exercise for the determination of $^{237}$Np using $^{242}$Pu as the yield tracer, which is the most critical stage of the proposed procedure, was carried using a reference seawater sample with a known $^{237}$Np activity (i.e. IAEA-443) that is measurable by AS (see section 2.5.3). The obtained results for the 3 studied samples are shown in Table 4. The average radiochemical yields were (69 ± 2)% and (70 ± 5)% for Np and Pu, respectively. Although these recoveries are slightly lower than the obtained ones during the testing stage with Monaco seawater (section 3.3), it is once more proved the same chemical behavior of Np and Pu in every processed sample. Two different estimations of the $^{237}$Np activity concentration were obtained in this case (section 2.5.3): i) from the $^{239}$Np added as a spike, and ii) using the $^{238}$Pu present in the original sample as the yield tracer. The obtained average result from the first approach agrees with the one from the second and both match the certified value within errors, which validates the proposed procedure.

**4.5 AMS results: IAEA-418**

In order to test the performance of the procedure for the AMS analysis of $^{239,240}$Pu + $^{237}$Np, and $^{236}$U in seawater samples, two aliquots of the IAEA-418 reference seawater sample were processed. The study of the IAEA-418 is interesting because corresponds to surface seawater collected at DYFAMED site in 2001 (north-western Mediterranean Sea), where results on $^{239+240}$Pu and $^{237}$Np activities have been also reported. Those values can be considered for comparison purposes [18,42]. On the other hand, $^{236}$U and $^{240}$Pu/$^{239}$Pu atom ratios have been obtained for the first time in this specific reference material in this work.

It is important to notice that, following the proposed procedure, in the final purified Np+Pu fractions traces of $^{238}$U at the 0.1 µg level or lower might be present (estimations from different tests performed by AS). During the AMS analysis on the 1 MV CNA system, the related $^{238}$U background effects can be corrected if a fine tuning is applied, as detailed in [38]. This might not be the case in conventional MS instruments, where the abundance sensitivity (i.e. mass suppression factor between neighboring masses) that can be achieved is usually much worse. A further adaptation of the method might be necessary in those cases.

The final values for the IAEA-418 are shown in Table 5. The results from the two aliquots are in agreement for all the studied radionuclides. The uncertainties can be considered acceptable (relative uncertainties ranging from 9% to 17% ) taking to account that only 1 L samples were processed for availability reasons. $^{239+240}$Pu obtained activities agree with the value reported in [42], (14 ± 2) µBq·L$^{-1}$, and $^{237}$Np results are also in agreement with the previous one, i.e. (0.22 ± 0.07) µBq·kg$^{-1}$ or (22 ± 7) at·kg$^{-1}$ [18]. It is important to remark that, in contrast with the 1 L aliquots processed in this work, 60 L samples were used in the case of the results reported in the previous works [18,42]. $^{240}$Pu/$^{239}$Pu atom ratios are in agreement with the expected value for global fallout (i.e. 0.18 ± 0.02 for the North Equatorial region [7]). Finally, $^{237}$Np/$^{236}$U ratios were lower than the expected one for global fallout in this area (i.e. 1.7-2.5 [7,8]). This could indicate the presence of an excess of $^{236}$U coming from a local source, in agreement with the reported studies of $^{236}$U in the Mediterranean Sea [14,53].

## 5. Conclusions

A new radiochemical method for a rapid sequential extraction of U, Np and Pu from small-volume seawater samples (<10 L) is presented. The method is optimized for the final $^{236}$U, $^{237}$Np, $^{239}$Pu and $^{240}$Pu determinations by AMS and is based on the use of $^{242}$Pu as tracer for both $^{237}$Np and Pu isotopes. High radiochemical yields (>80%) were obtained for the three elements and the ratios between the Np and Pu radiochemical yields were, on average 0.98 ± 0.04. Besides, the reliability of the use of $^{242}$Pu as $^{237}$Np tracer was validated by analysing the reference seawater sample IAEA-443 by radiometrics techniques. Because of the very low concentrations that can be expected from seawater samples, laboratory background was optimized by an in-depth study of the different reagents. This study reveals Fe(SO)$_4$ comercial salts as a possible source of $^{236}$U background. The method is finally apply for the study of the reference sampe IAEA-418 (Mediterranean seawater sample). Results were consistent for $^{236}$U and $^{240}$Pu/$^{239}$Pu, which were reported for the first time in this reference material, and $^{237}$Np and $^{239+240}$Pu activities agreed with the reported values. Therefore, the method has been sucesfully validate and inplemented for AMS analys at the CNA, opening new possibilities for the determination of actinides in oceanographic campaigns, where the simplification of the procedures and the optimization of the sample volumes are key points. Furthermore, the possibility of studying $^{237}$Np by using a simple radiochemical method allows us to increase the very limited $^{237}$Np dataset available from oceanographic samples.

## Acknowledgments

This work has been financed from the project FIS2015-69673-P, provided by the Spanish Ministry of Economy. This work was partially funded by Fundación Cámara Sevilla through a Grant for Graduate Studies. The IAEA is grateful to the Government of the Principality of Monaco for the support provided to its Environment Laboratories.

## Data availability

The data that support the findings of this study are openly available at the following URL: https://github.com/AMS-CNA/Method-U-Np-Pu-Seawater

# References

[1]  A. Sakaguchi, K. Kawai, P. Steier, F. Quinto, K. Mino, J. Tomita, M. Hoshi, N. Whitehead, M. Yamamoto, First results on 236U levels in global fallout, Sci. Total Environ. 407 (2009) 4238–4242. doi:10.1016/j.scitotenv.2009.01.058.

[2]  S.R. Winkler, P. Steier, J. Carilli, Bomb fall-out 236U as a global oceanic tracer using an annually resolved coral core, Earth Planet. Sci. Lett. 359–360 (2012) 124–130. doi:10.1016/J.EPSL.2012.10.004.

[3]  T.M. Beasley, J.M. Kelley, T.C. Maiti, L.A. Bond, 237Np239Pu atom ratios in integrated global fallout: a reassessment of the production of 237Np, J. Environ. Radioact. 38 (1998) 133–146. doi:10.1016/S0265-931X(97)00033-7.

[4]  T. Beasley, L.W. Cooper, J.M. Grebmeier, K. Aagaard, J.M. Kelley, L.R. Kilius, 237Np/129I atom ratios in the Arctic Ocean: Has 237Np from Western European and Russian fuel reprocessing facilities entered the Arctic Ocean?, J. Environ. Radioact. 39 (1998) 255–277. doi:10.1016/S0265-931X(97)00059-3.

[5]  M. Christl, N. Casacuberta, C. Vockenhuber, C. Elsaasser, Pascal Bailly du Bois, J. Herrmann, H.-A. Synal, Reconstruction of the 236U input function for the Northeast Atlantic Ocean: Implications for 129I/236U and 236U/238U-based tracer ages, J. Geophys. Res. Ocean. 120 (2015) 7282–7299. doi:doi:10.1002/ 2015JC011116.

[6]  N. Casacuberta, P. Masqué, G. Henderson, M. Rutgers van-der-Loeff, D. Bauch, C. Vockenhuber, A. Daraoui, C. Walther, H.-A.A. Synal, M. Christl, First 236U data from the Arctic Ocean and use of 236U/238U and 129I/236U as a new dual tracer, Earth Planet. Sci. Lett. 440 (2016) 127–134. doi:http://dx.doi.org/10.1016/j.epsl.2016.02.020.

[7]  J.M. Kelley, L.A. Bond, T.M. Beasley, Global distribution of Pu isotopes and 237Np, Sci. Total Environ. 237–238 (1999) 483–500. doi:http://dx.doi.org/10.1016/S0048-9697(99)00160-6.

[8]  E. Chamizo, M. López-Lora, M. Villa, N. Casacuberta, J.M. López-Gutiérrez, M.K. Pham, Analysis of 236U and plutonium isotopes, 239,240Pu, on the 1 MV AMS system at the Centro Nacional de Aceleradores, as a potential tool in oceanography, Nucl. Instruments Methods Phys. Res. Sect. B Beam Interact. with Mater. Atoms. 361 (2015) 535–540. doi:10.1016/j.nimb.2015.02.066.

[9]  P.P. Povinec, C. Badie, A. Baeza, Et al., Certified reference material for radionuclides in seawater IAEA-381 (Irish Sea Water), J. Radioanal. Nucl. Chem. 251 (2002) 369–374. doi:10.1023/A:1014861620713.

[10] U.N.S.C. on the E. of A. Radiation, Exposures and effects of the Chernobyl accident, in: UNSCEAR 2000 Rep. Vol. II, 2000: p. 115.

[11] A. Sakaguchi, A. Kadokura, P. Steier, Y. Takahashi, K. Shizuma, M. Hoshi, T. Nakakuki, M. Yamamoto, Uranium-236 as a new oceanic tracer: A first depth profile in the Japan Sea and comparison with caesium-137, Earth Planet. Sci. Lett. 333–334 (2012) 165–170. doi:10.1016/j.epsl.2012.04.004.

[12] N. Casacuberta, M. Christl, J. Lachner, M.R. van der Loeff, P. Masqué, H.A. Synal, A first

transect of 236U in the North Atlantic Ocean, Geochim. Cosmochim. Acta. 133 (2014) 34–46. doi:http://dx.doi.org/10.1016/j.gca.2014.02.012.

[13] M. Christl, J. Lachner, C. Vockenhuber, I. Goroncy, J. Herrmann, H.-A. Synal, First data of Uranium-236 in the North Sea, Nucl. Instruments Methods Phys. Res. Sect. B Beam Interact. with Mater. Atoms. 294 (2013) 530–536. doi:10.1016/j.nimb.2012.07.043.

[14] E. Chamizo, M. López-Lora, M. Bressac, I. Levy, M.K. Pham, Excess of 236 U in the northwest Mediterranean Sea, Sci. Total Environ. 565 (2016) 767–776. doi:10.1016/j.scitotenv.2016.04.142.

[15] R. Eigl, P. Steier, K. Sakata, A. Sakaguchi, Vertical distribution of 236U in the North Pacific Ocean, J. Environ. Radioact. 169–170 (2017) 70–78. doi:10.1016/J.JENVRAD.2016.12.010.

[16] R.J. Pentreath, B.R. Harvey, The presence of 237Np in the Irish Sea, Mar. Ecol. Prog. Ser. 6 (1981) 243–247.

[17] P. Lindahl, P. Roos, E. Holm, H. Dahlgaard, Studies of Np and Pu in the marine environment of SwedisheDanish waters and the North Atlantic Ocean, J. Environ. Radioact. 82 (2005) 285–301. doi:10.1016/j.jenvrad.2005.01.011.

[18] M. Bressac, I. Levy, E. Chamizo, J.J. La Rosa, P.P. Povinec, J. Gastaud, B. Oregioni, Temporal evolution of 137 Cs, 237 Np, and 239+240 Pu and estimated vertical 239+240 Pu export in the northwestern Mediterranean Sea, Sci. Total Environ. 595 (2017) 178–190. doi:10.1016/j.scitotenv.2017.03.137.

[19] M. Castrillejo Iridoy, Sources and distribution of artificial radionuclides in the oceans: from Fukushima to the Mediterranean Sea, n.d. http://www.tdx.cat/bitstream/handle/10803/457438/mci1de1.pdf?sequence=1 (accessed December 28, 2017).

[20] P.P. Parekh, T.M. Semkow, M.A. Torres, D.K. Haines, J.M. Cooper, P.M. Rosenberg, M.E. Kitto, Radioactivity in Trinitite six decades later, J. Environ. Radioact. 85 (2006) 103–120. doi:10.1016/J.JENVRAD.2005.01.017.

[21] T. Bisinger, S. Hippler, R. Michel, L. Wacker, H.-A. Synal, Determination of plutonium from different sources in environmental samples using alpha-spectrometry and AMS, Nucl. Inst. Methods Phys. Res. B. 268 (2009) 1269–1272. doi:10.1016/j.nimb.2009.10.150.

[22] M. Villa-alfageme, E. Chamizo, T.C. Kenna, M. López-Lora, N. Casacuberta, P. Masqué, M. Christl, Distribution of 236U in the GEOTRACES EPZT and its use as a water mass tracer (In press), Chem. Geol. (n.d.).

[23] K. Hirose, M. Aoyama, M. Fukasawa, C.S. Kim, K. Komura, P.P. Povinec, J.A. Sanchez-Cabeza, Plutonium and 137Cs in surface water of the South Pacific Ocean, Sci. Total Environ. 381 (2007) 243–255. doi:10.1016/J.SCITOTENV.2007.03.022.

[24] J. Zheng, M. Yamada, Inductively coupled plasma-sector field mass spectrometry with a high-efficiency sample introduction system for the determination of Pu isotopes in settling particles at femtogram levels, Talanta. 69 (2006) 1246–1253. doi:10.1016/j.talanta.2005.12.047.

[25] V.N. Epov, K. Benkhedda, R.J. Cornett, R.D. Evans, Rapid determination of plutonium in urine using flow injection on-line preconcentration and inductively coupled plasma mass spectrometry, J. Anal. At. Spectrom. 20 (2005) 424. doi:10.1039/b501218j.

[26] C.S. Kim, C.K. Kim, K.J. Lee, Simultaneous analysis of 237Np and Pu isotopes in environmental samples by ICP-SF-MS coupled with automated sequential injection system, J. Anal. At. Spectrom. 19 (2004) 743. doi:10.1039/b400034j.

[27] L.K. Fifield, Accelerator mass spectrometry of the actinides, Quat. Geochronol. 3 (2008) 276–290. doi:10.1016/J.QUAGEO.2007.10.003.

[28] I. Levy, Détermination des radionucléides anthropiques plutonium, neptunium, américium, césium et strontium dans l'environnement marin Méditerranéen., 2004.

[29] M. López-Lora, E. Chamizo, M. Villa-Alfageme, S. Hurtado-Bermúdez, N. Casacuberta, M. García-León, Isolation of $^{236}$U and $^{239,240}$Pu from seawater samples and its determination by Accelerator Mass Spectrometry, Talanta. 178 (2018). doi:10.1016/j.talanta.2017.09.026.

[30] J. Qiao, X. Hou, P. Steier, R. Golser, Sequential Injection Method for Rapid and Simultaneous Determination of $^{236}$U, $^{237}$Np, and Pu Isotopes in Seawater, Anal. Chem. 85 (2013) 11026–11033. doi:10.1021/ac402673p.

[31] S.L. Maxwell, B.K. Culligan, D.R. Mcalister, Rapid determination of actinides in seawater samples, (2014) 1175–1189. doi:10.1007/s10967-014-3079-0.

[32] J. La Rosa, L. Gastaud, L. Lagan, S.-H. Lee, I. Levy-Palomo, P.P. Povinec, E. Wyse, Recent developments in the analysis of transuranics (Np, Pu, Am) in seawater, J. Radioanal. Nucl. Chem. 236 (2005) 9.

[33] P. Thakur, G.P. Mulholland, Determination of 237Np in environmental and nuclear samples: A review of the analytical method, Appl. Radiat. Isot. 70 (2012) 1747–1778. doi:10.1016/j.apradiso.2012.02.115.

[34] E.P. Horwitz, M.L. Dietz, R. Chiarizia, H. Diamond, S.L. Maxwell, M.R. Nelson, Separation and preconcentration of actinides by extraction chromatography using a supported liquid anion exchanger: application to the characterization of high-level nuclear waste solutions, Anal. Chim. Acta. 310 (1995) 63–78. doi:10.1016/0003-2670(95)00144-O.

[35] M.J. Keith-Roach, J.P. Day, F.R. Livens, L.K. Fifield, Measurement of 237Np in environmental water samples by accelerator mass spectrometry, Analyst. 126 (2001) 58–61. doi:10.1039/b007493o.

[36] Q. Chen, H. Dahlgaard, S.P. Nielsen, A. Aarkrog, Pu as tracer for simultaneous determination of 237 Np and 239,240 Pu in environmental samples, J. Radioanal. Nucl. Chem. 253 (2002) 451–458. https://link.springer.com/content/pdf/10.1023%2FA%3A1020429805654.pdf (accessed December 22, 2017).

[37] S.L. Maxwell, B.A. Culligan, V.D. Jones, S.T. Nichols, G.W. Noyes, Rapid determination of 237Np and Pu isotopes in water by inductively-coupled plasma mass spectrometry and alpha spectrometry, J. Radioanal. Nucl. Chem. 287 (2011) 223–230. doi:10.1007/s10967-

010-0825-9.

[38] M. López-Lora, E. Chamizo, Acelerator Mass Spectrometry of 237Np 239Pu and 240Pu for environmental aplications at the Centro Nacional de Aceleradores (CNA) (Submitted), Nucl. Instruments Methods Phys. Res. Sect. B Beam Interact. with Mater. Atoms. (n.d.).

[39] J. La Rosa, I. Outola, E. Crawford, S. Nour, H. Kurosaki, K. Inn, Radiochemical measurement of 237Np in a solution of mixed radionuclides: Experiences in chemical separation and alpha-spectrometry, J. Radioanal. Nucl. Chem. 277 (2008) 11–18. doi:10.1007/s10967-008-0702-y.

[40] CERTIFIED REFERENCE MATERIAL IAEA-443 NATURAL AND ARTFICIAL RADIONUCLIDES IN SEA WATER FROM THE IRISH SEA, (n.d.). https://nucleus.iaea.org/rpst/referenceproducts/referencematerials/radionuclides/IAEA-443/RS_IAEA_443_final.pdf (accessed December 13, 2017).

[41] IAEA-418 IODINE-129 IN MEDITERRANEAN SEA WATER, (n.d.). https://nucleus.iaea.org/rpst/ReferenceProducts/ReferenceMaterials/Radionuclides/IAEA-418/RS_IAEA_418.pdf (accessed December 13, 2017).

[42] S.-H. Lee, J.J. La Rosa, I. Levy-Palomo, B. Oregioni, M.K. Pham, P.P. Povinec, E. Wyse, Recent inputs and budgets of 90Sr, 137Cs, 239,240Pu and 241Am in the northwest Mediterranean Sea, Deep Sea Res. Part II Top. Stud. Oceanogr. 50 (2003) 2817–2834. doi:http://dx.doi.org/10.1016/S0967-0645(03)00144-9.

[43] I. Levy, P.P. Povinec, M. Aoyama, K. Hirose, J.A. Sanchez-Cabeza, J.-F. Comanducci, J. Gastaud, M. Eriksson, Y. Hamajima, C.S. Kim, K. Komura, I. Osvath, P. Roos, S.A. Yim, Marine anthropogenic radiotracers in the Southern Hemisphere: New sampling and analytical strategies, Prog. Oceanogr. 89 (2011) 120–133. doi:10.1016/j.pocean.2010.12.012.

[44] E.C. Calvo, F.J. Santos, J.M. López-Gutiérrez, S. Padilla, M. García-León, J. Heinemeier, C. Schnabel, G. Scognamiglio, Status report of the 1 MV AMS facility at the Centro Nacional de Aceleradores, Nucl. Instruments Methods Phys. Res. Sect. B Beam Interact. with Mater. Atoms. 361 (2015) 13–19. doi:10.1016/j.nimb.2015.02.022.

[45] G. Scognamiglio, E. Chamizo, J.M. López-Gutiérrez, A.M. Müller, S. Padilla, F.J. Santos, M. López-Lora, C. Vivo-Vilches, M. García-León, Recent developments of the 1 MV AMS facility at the Centro Nacional de Aceleradores, Nucl. Instruments Methods Phys. Res. Sect. B Beam Interact. with Mater. Atoms. 375 (2016) 17–25. doi:10.1016/j.nimb.2016.03.033.

[46] E. Chamizo, M. Christl, L.K. Fifield, Measurement of 236U on the 1 MV AMS system at the Centro Nacional de Aceleradores (CNA), Nucl. Instruments Methods Phys. Res. Sect. B Beam Interact. with Mater. Atoms. 358 (2015) 45–51. doi:http://dx.doi.org/10.1016/j.nimb.2015.05.008.

[47] E. Chamizo, M. López-Lora, Accelerator mass spectrometry of 236 U with He stripping at the Centro Nacional de Aceleradores, Nucl. Instruments Methods Phys. Res. Sect. B Beam Interact. with Mater. Atoms. (2018). doi:10.1016/j.nimb.2018.04.020.

[48] E.P. Horwitz, M.L. Dietz, R. Chiarizia, H. Diamond, A.M. Essling, D. Graczyk, Separation and preconcentration of uranium from acidic media by extraction chromatography, Anal. Chim.

Acta. 266 (1992) 25–37. doi:10.1016/0003-2670(92)85276-C.

[49] L. Hallstadius, A method for the electrodeposition of actinides, Nucl. Instruments Methods Phys. Res. 223 (1984) 266–267. doi:http://dx.doi.org/10.1016/0167-5087(84)90659-8.

[50] J. Qiao, X. Hou, P. Steier, S. Nielsen, R. Golser, Method for $^{236}$U Determination in Seawater Using Flow Injection Extraction Chromatography and Accelerator Mass Spectrometry, Anal. Chem. 87 (2015) 7411–7417. doi:10.1021/acs.analchem.5b01608.

[51] C.W. Sill, F.D. Hindman, J.I. Anderson, Simultaneous determination of alpha-emitting nuclides of radium through californium in large environmental and biological samples, Anal. Chem. 51 (1979) 1307–1314. doi:10.1021/ac50044a043.

[52] M. Mendes, J. Aupiais, C. Jutier, F. Pointurier, Determination of weight distribution ratios of Pa(V) and Np(V) with some extraction chromatography resins and the AG1-X8 resin, Anal. Chim. Acta. 780 (2013) 110–116. doi:10.1016/J.ACA.2013.04.019.

[53] M. Castrillejo, N. Casacuberta, M. Chirstl, C. Vockenhuber, P. Masqué, J. García-Orellana, Mapping of 236U and 129I in the Mediterranean Sea, Annu. Rep. 2014, Ion Beam Physics, ETH Zurich. (2014).

# Tables

*Table 1- U and Np co-precipitation yields obtained from 10 L seawater samples increasing the pH to different values for the formation of the $Fe(OH)_2$ precipitate. In samples A, B and C, Np yields were quantified separately from a first precipitate ($1^{st}$ ppt) and a second one obtained from the supernatant ($2^{nd}$ ppt); both precipitated were finally combined (final). U results from these samples were obtained by the analysis of the combined precipitates. In samples D, E and F, only one precipitate was recovered. The average results (Avg.) together with the Standard Deviation (SD) are shown for each set of samples.*

| Sample (10 L seawater) | pH | | Co-precipitation yield (%) | | | |
|---|---|---|---|---|---|---|
| | | | U | Np ($1^{st}$ ppt) | Np ($2^{nd}$ ppt) | Np (final) |
| A | 8 - 8.5 | | 80 ± 10 | 69 ± 2 | 16.2 ± 0.6 | 86 ± 2 |
| B | 8 - 8.5 | | 25 ± 4 | 69 ± 3 | 11.0 ± 0.7 | 80 ± 3 |
| C | 8 - 8.5 | | 70 ± 10 | 67 ± 2 | 12.5 ± 0.6 | 80 ± 2 |
| | | Avg. | 60 | 68 | 13 | 82 |
| | | SD | 30 | 1 | 3 | 3 |
| D | 9 - 9.5 | | 83 ± 9 | 96 ± 2 | - | 96 ± 2 |
| E | 9 - 9.5 | | 110 ± 10 | 88 ± 2 | - | 88 ± 2 |
| F | 9 - 9.5 | | 100 ± 10 | 94 ± 4 | - | 94 ± 4 |
| | | Avg. | 98 | 93 | - | 93 |
| | | SD | 9 | 4 | - | 4 |

*Table 2- Results obtained from two set of column-blank samples processed with different oxidation state adjustments in order to study the radiochemical yield during the extraction chromatography step. The average results (Avg.) together with the Standard Deviation (SD) are shown for each set of samples. In the last column the ratios between the Np and Pu yields are calculated.*

| Oxidation state adjustment | Sample | | Extraction Chromatography Yield (%) | | | Np/Pu (yield ratios) |
|---|---|---|---|---|---|---|
| | | | U | Np | Pu | |
| - Dissolution in 17 mL of 1 M $HNO_3$ - 200 mg of Mohr's Salt - 2.75 mL $HNO_3$ (70%) | Blk-1 | | 100 ± 10 | 90 ± 4 | 82 ± 8 | 1.1 ± 0.1 |
| | Blk-2 | | 110 ± 10 | 92 ± 5 | 91 ± 9 | 1.0 ± 0.1 |
| | Blk-3 | | 100 ± 10 | 92 ± 5 | 93 ± 9 | 1.0 ± 0.1 |
| | | Avg. | 103 | 91.2 | 89 | 1.03 |
| | | SD | 6 | 1.1 | 6 | 0.06 |
| - Dissolution in 20 mL of 1 M $HNO_3$ - 200 mg of Mohr's Salt - 3.25 mL $HNO_3$ (70%) | Blk-4 | | 93 ± 9 | 94 ± 5 | 100 ± 10 | 0.9 ± 0.1 |
| | Blk-5 | | 66 ± 7 | 98 ± 4 | 97 ± 9 | 1.0 ± 0.1 |
| | Blk-6 | | 100 ± 10 | 95 ± 3 | 100 ± 10 | 0.9 ± 0.1 |
| | | Avg. | 90 | 96 | 99 | 0.9 |
| | | SD | 20 | 2 | 2 | 0.1 |

*Table 3- Radiochemical yields obtained from a set of seawater samples processed following the proposed procedure. In the last column the ratios between the Np and Pu yields are calculated. The average results (Avg.) together with the Standard Deviation (SD) are shown. U and Pu yields were quantified as it is described in Section 2.5.*

| Sample | Seawater aliquot (L) | Radiochemial Yield (%) | | | Np/Pu (yield ratios) |
|---|---|---|---|---|---|
| | | U | Np | Pu | |
| 1 | 10 | 82 ± 9 | 81 ± 3 | 81 ± 7 | 1.00 ± 0.09 |
| 2 | 10 | 90 ± 10 | 60 ± 3 | 61 ± 6 | 1.0 ± 0.1 |
| 3 | 10 | 85 ± 9 | 83 ± 4 | 88 ± 9 | 0.9 ± 0.1 |
| 4 | 2.5 | 100 ± 10 | 83 ± 3 | 81 ± 8 | 1.0 ± 0.1 |
| 5 | 2.5 | 100 ± 10 | 88 ± 3 | 87 ± 9 | 1.0 ± 0.1 |
| 6 | 2.5 | 100 ± 10 | 90 ± 3 | 94 ± 9 | 1.0 ± 0.1 |
| | Avg. | 93 | 80 | 80 | 0.98 |
| | SD | 8 | 10 | 10 | 0.04 |

*Table 4- Obtained results for the IAEA-443 (Irish Sea Sample). Np and Pu recoveries are shown together with the measured activities of $^{237}$Np obtained by using $^{239}$Np and $^{238}$Pu as tracers. The average results (Avg.) and the Standard Deviation (SD) are calculated in each case. The expected $^{237}$Np activity, reported in [9], is also shown.*

| Sample | Radiochemical yield (%) | | $^{237}$Np (mBq kg$^{-1}$) | | Expected value |
|---|---|---|---|---|---|
| | | | Measured value | | |
| | Np | Pu | Tracer: $^{239}$Np | Tracer: $^{238}$Pu | |
| IAEA-443-a | 68 ± 3 | 65 ± 8 | 10 ± 1 | 10 ± 1 | |
| IAEA-443-b | 69 ± 3 | 74 ± 8 | 9 ± 1 | 9 ± 1 | |
| IAEA-443-c | 71 ± 2 | 70 ± 10 | 9 ± 1 | 9 ± 1 | |
| Avg. | 69 | 70 | 9.3 | 9.3 | 8.7 ± 0.5 |
| SD | 2 | 5 | 0.6 | 0.6 | |

*Table 5- Results from 2 samples of 1 L aliquots of the IAEA-418 (Mediterranean Sea water). Samples were processed following the proposed method and analyzed by AMS at the CNA. Concentrations of the different radionuclides of interest ($^{236}$U, $^{237}$Np, $^{239}$Pu and $^{240}$Pu) are shown together with the $^{236}$U/$^{238}$U and $^{240}$Pu/$^{239}$Pu atom ratios.*

| Sample | Concentration (10$^6$ atoms kg$^{-1}$) | | | | $^{236}$U/$^{238}$U (10$^{-9}$) | $^{240}$Pu/$^{239}$Pu | $^{237}$Np/$^{236}$U | $^{239+240}$Pu (µBq kg$^{-1}$) |
|---|---|---|---|---|---|---|---|---|
| | $^{236}$U | $^{237}$Np | $^{239}$Pu | $^{240}$Pu | | | | |
| 418-a | 26 ± 3 | 16 ± 3 | 10.3 ± 0.9 | 2.1 ± 0.2 | 3.2 ± 0.3 | 0.20 ± 0.03 | 0.6 ± 0.2 | 17 ± 1 |
| 418-b | 27 ± 3 | 21 ± 3 | 11 ± 1 | 1.7 ± 0.2 | 3.4 ± 0.4 | 0.16 ± 0.02 | 0.8 ± 0.2 | 16 ± 1 |

# Figures

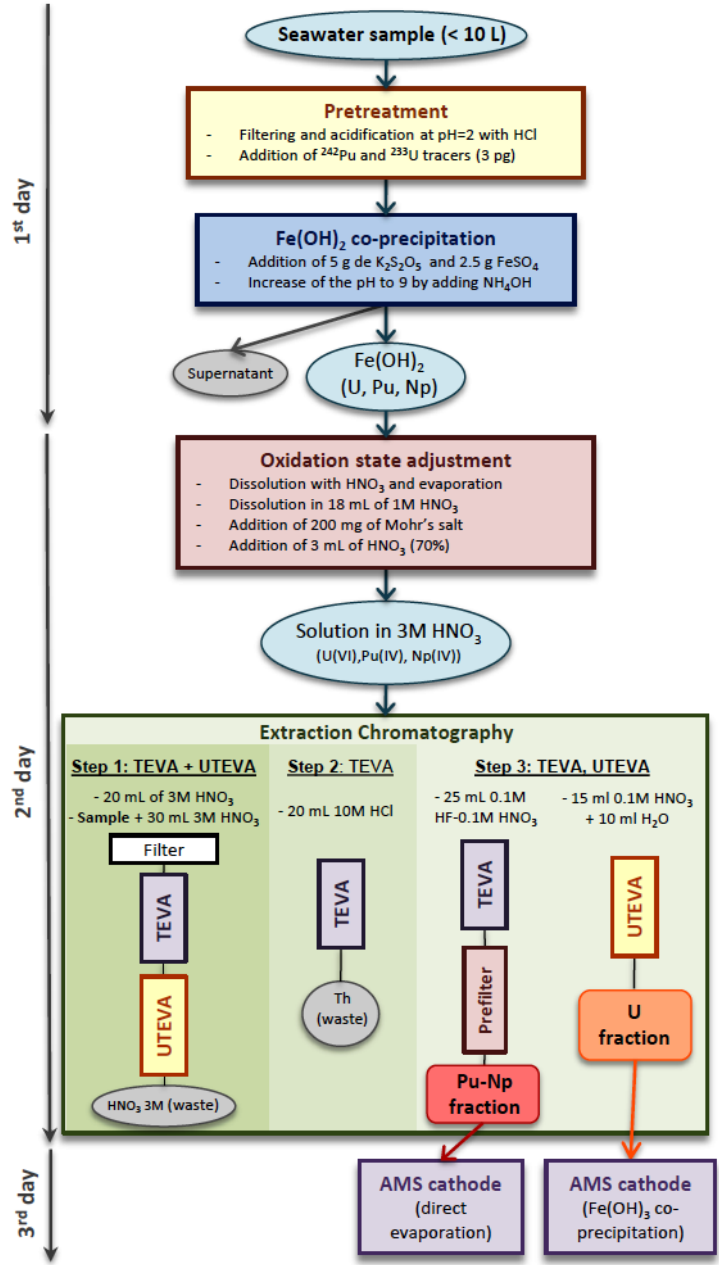

***Fig. 1**- Flow-diagram of the procedure proposed in this paper for the sequential extraction of U, Np and Pu from seawater.*

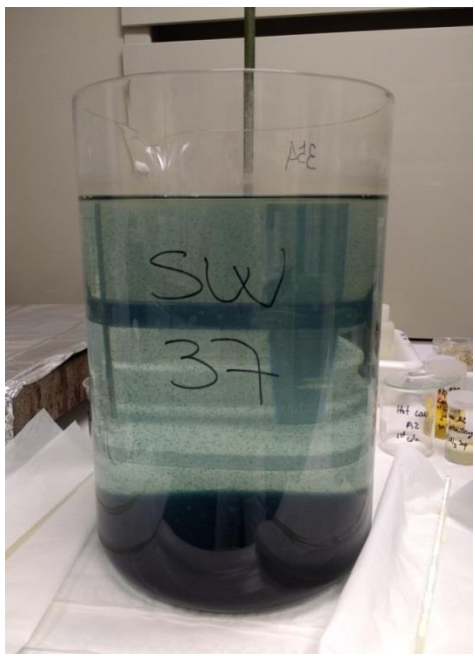

***Fig. 2-*** *Fe(OH)$_2$ precipitation for a 10L seawater sample.*

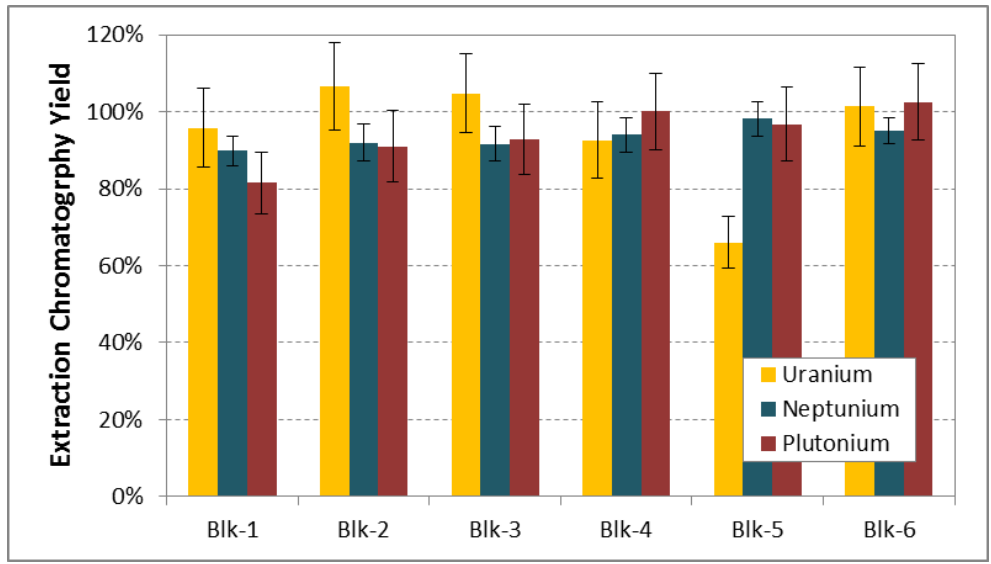

***Fig. 3-*** *Radiochemical yield during the extraction chromatography step obtained from a set of 6 column-blank samples spiked with $^{232}$U, $^{239}$Np and $^{242}$Pu. Different oxidations state adjustments were applied for the first group of samples (Blk-1, Blk-2, Blk-3) and for the second one (Blk-4, Blk-5, Blk-6). Values showed in Table 2.*

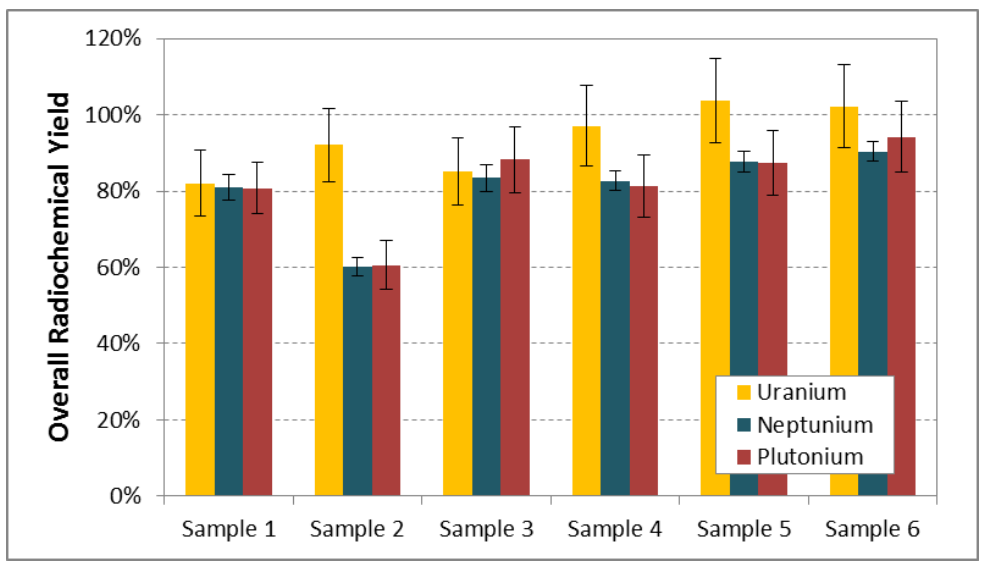

***Fig. 4-*** *Radiochemical yields obtained from seawater samples aliquots of 10 L (samples 1, 2 and 3) and 2.5 L (samples 4, 5 and 6) following the proposed method described in the section 2.3.*